\documentclass[conference]{IEEEtran} 

\usepackage{multirow}
\usepackage{balance}
\usepackage{color}
\usepackage{epsfig}

\def\IEEEbibitemsep{3pt} 

\ifCLASSOPTIONcompsoc
 \usepackage[caption=false,font=normalsize,labelfont=sf,textfont=sf]{subfig}
\else
 \usepackage[caption=false,font=footnotesize]{subfig}
\fi

\newcommand{\wzk}[1]{{\color{black}#1}}
\newcommand{\blue}[1]{{\color{blue}#1}}

\graphicspath{{./figure/}}

\newcommand{\FrameworkName}{{Shuhai}}

\author{
\IEEEauthorblockN{Zeke Wang, Hongjing Huang, Jie Zhang}
 \IEEEauthorblockA{
	Collaborative Innovation Center of Artificial Intelligence\\ Zhejiang University, China\\
	Email: \{wangzeke, 21515069, carlzhang4\}@zju.edu.cn }

	\and
  
\IEEEauthorblockN{Gustavo Alonso }
 \IEEEauthorblockA{Systems Group\\ ETH Zurich, Switzerland\\
	Email: alonso@inf.ethz.ch}
}

\begin{document}

\title{Benchmarking High Bandwidth Memory on FPGAs}
\maketitle

\vspace{0.5ex}
\begin{abstract}
 FPGAs are starting to be enhanced with High Bandwidth Memory (HBM) as a way to reduce the memory bandwidth bottleneck encountered in some applications and to give the FPGA more capacity to deal with application state. However, the performance characteristics of HBM are still not well specified, especially in the context of FPGAs. In this paper, we bridge the gap between nominal specifications and actual performance by benchmarking HBM on a state-of-the-art FPGA, i.e., a Xilinx Alveo U280 featuring a two-stack HBM subsystem.   
To this end, we propose \FrameworkName{}, a benchmarking tool that allows us to demystify all the underlying details of HBM on an FPGA. FPGA-based benchmarking should also provide a more accurate picture of HBM than doing so on CPUs/GPUs, since CPUs/GPUs are noisier systems due to their complex control logic and cache hierarchy. 
Since the memory itself is complex, leveraging custom hardware logic to benchmark inside an FPGA provides more details as well as accurate and deterministic measurements. We observe that 1) HBM is able to provide up to 425 GB/s memory bandwidth, and 2) how HBM is used has a significant impact on performance, which in turn demonstrates the importance of unveiling the performance characteristics of HBM so as to select the best approach. 
\FrameworkName{} can be easily generalized to other FPGA boards or other generations of memory, e.g., HBM3, and DDR3. We will make \FrameworkName{} open-source, benefiting the community. 

\end{abstract}

\vspace{0.5ex}
\section{Introduction}


The computational capacity of modern computing system continues increasing due to the constant improvements on CMOS technology, typically by instantiating more cores within the same area and/or by adding extra functionality to the cores (AVX, SGX, etc.). In contrast, the bandwidth capability of DRAM memory has only slowly improved over many generations. As a result, the gap between memory and processor speed keeps growing and is being exacerbated by  multicore designs due to the concurrent access. To bridge the memory bandwidth gap, semiconductor memory companies such as Samsung\footnote{https://www.samsung.com/semiconductor/dram/hbm2/} have released a few memory variants, e.g., Hybrid Memory Cube (HMC) and High Bandwidth Memory (HBM), as a way to provide significantly higher memory ba ndwidth. For example, the state-of-the-art Nvidia GPU V100 features 32 GB HBM2 (the second generation HBM) to provide up to 900 GB/s memory bandwidth for its thousands of computing cores.\footnote{{https://www.nvidia.com/en-us/data-center/v100/}}

Compared with a GPU of the same generation, FPGAs used to have an order of magnitude lower memory bandwidth since FPGAs typically feature up to 2 DRAM memory channels, each of which has up to 19.2 GB/s memory bandwidth on our tested FPGA board Alevo U280~\cite{u280}.\footnote{https://www.xilinx.com/products/boards-and-kits/alveo/u280.html} As a result, an FPGA-based solution using DRAM could not compete with a GPU for bandwidth-critical applications. Consequently, \wzk{FPGA vendors like Xilinx~\cite{u280} have started to introduce HBM\footnote{In the following, we use HBM which refers to HBM2 in the context of Xilinx FPGAs, as Xilinx FPGAs feature two HBM2 stacks.} in their FPGA boards as a way to remain competitive on those same applications}. HBM has the potential to be a game-changing feature by allowing FPGAs to provide significantly higher performance for memory- and compute-bound applications like database engines~\cite{partition_fpl15} or deep learning inference~\cite{cnn_fpt17}. It can also support applications in keeping more state within the FPGA without the significant performance penalties seen today as soon as DRAM is involved. 


Despite the potential of HBM to bridge the bandwidth gap, there are still obstacles to leveraging HBM on the FPGA. 
First, the performance characteristics of HBM are often unknown to developers, especially to FPGA programmers. Even though an HBM stack consists of a few traditional DRAM dies and a logic die, the performance characteristics of HBM are significantly different than those of, e.g., DDR4. Second, Xilinx's HBM subsystem~\cite{hbm_ip} introduces new features like a \emph{switch} inside its HBM memory controller. The performance characteristics of the switch are also unclear to the FPGA programmer due to the limited details exposed by Xilinx. These two issues can hamper the ability of FPGA developers to fully exploit the advantages of HBM on FPGAs.   

To this end, we present \FrameworkName{},\footnote{\FrameworkName{} is a pioneer of Chinese measurement standards, with which he measured the territory of China in the Xia dynasty. } a benchmarking tool that allows us to demystify all the underlying details of HBM. \FrameworkName{} adopts a software/hardware co-design approach to provide \emph{high-level insights} and \emph{ease of use} to developers or researchers interested in leveraging HBM. The high-level insights come from the first end-to-end analysis of the performance characteristic of typical \emph{memory access patterns}. The ease of use arises from the fact that \FrameworkName{} performs the majority of the benchmarking task without having to reconfigure the FPGA between parts of the benchmark. To our knowledge, \FrameworkName{} is the first platform to systematically benchmark HBM on an FPGA. We demonstrate the usefulness of \FrameworkName{} by identifying four important aspects on the usage of HBM-enhanced FPGAs:

\vspace{0.5ex}
\begin{itemize}
\item HBMs Provide Massive Memory Bandwidth.On the tested FPGA board Alveo U280, HBM provides up to 425 GB/s memory bandwidth, an order of magnitude more than using two traditional DDR4 channels on the same board. This is still half of what state-of-the-art GPUs obtain but it represents a significant leap forward for FPGAs.
\item The Address Mapping Policy is Critical to High Bandwidth. Different address mapping policies lead to an order of magnitude throughput differences when running a typical memory access pattern (i.e., sequential traversal) on HBM, indicating the importance of matching the address mapping policy to a particular application.
\item Latency of HBM is Much Higher than DDR4. The connection between HBM chips and the associated FPGA is done via serial I/O connection, introducing extra processing for parallel-to-serial-to-parallel conversion. For example, \FrameworkName{} identifies that the latency of HBM is 106.7 ns while the latency of DDR4 is 73.3 ns, when the memory transaction hits an open page (or row), indicating that we need more on-the-fly memory transactions, which are allowed on modern FPGAs/GPUs, to saturate HBM. 
\item FPGA Enables Accurate Benchmarking Numbers. We have implemented \FrameworkName{} on an FPGA with the benchmarking engine directly attaching to HBM modules, making it easier to reason about the performance numbers from HBM. In contrast, benchmarking memory performance on CPUs/GPUs makes it difficult to distinguish effects as, e.g., the cache introduces significant interference in the measurements. Therefore, we argue that our FPGA-based benchmarking approach is a better option when benchmarking memory, whether HBM or DDR.
\end{itemize}

\vspace{0.5ex}

\vspace{0.5ex}
\section{Background}
An HBM chip employs the latest development of IC packaging technologies, such as Through Silicon Via (TSV), stacked-DRAM, and 2.5D package~\cite{hbm_imw17, interposer_edaps15, new_memory_micro17, amd_fury_hbm_hcs15}. The basic structure of HBM consists of a base logic die at the bottom and 4 or 8 core DRAM dies stacked on top. All the dies are interconnected by TSVs.

\begin{figure}[t]
	\centering
	\includegraphics[width=8.6cm]{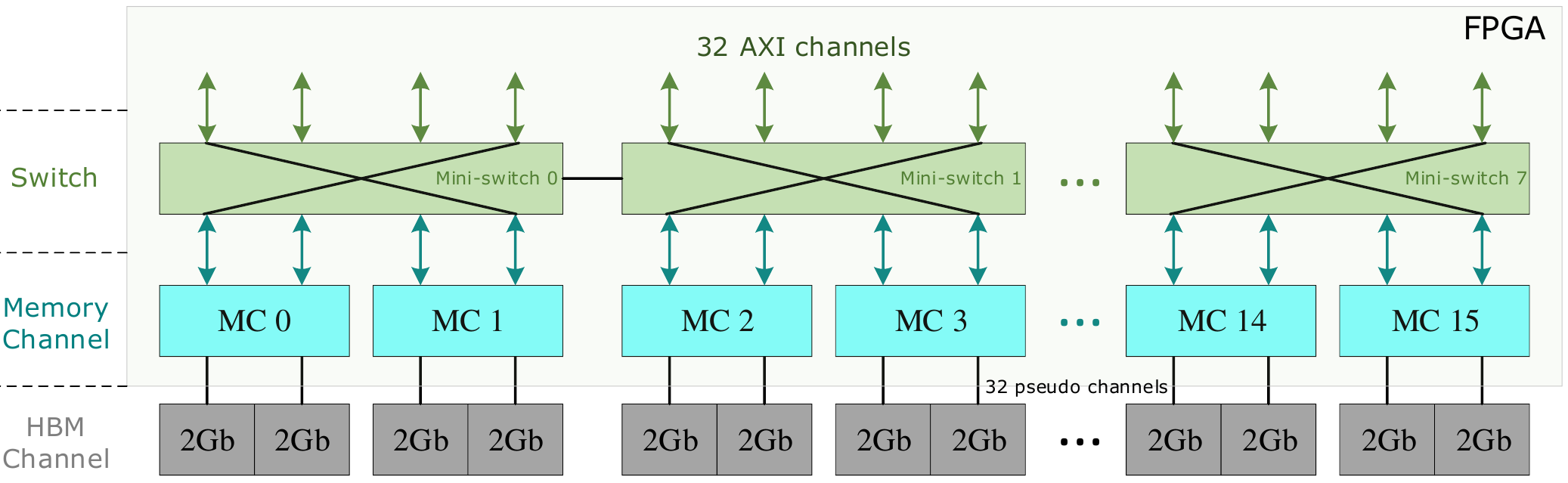}
	\caption{Architecture of Xilinx HBM subsystem }
	\vspace{-3ex}
	\label{P_xilinx_HBM}
\end{figure}

Xilinx integrates two \emph{HBM stacks} and an HBM controller inside the FPGA. Each HBM stack is divided into eight independent \emph{memory channels}, where each memory channel is further divided into two 64-bit \emph{pseudo channels}. A pseudo channel is only allowed to access its associated \emph{HBM channel} that has its own address region of memory, as shown in Figure~\ref{P_xilinx_HBM}. The Xilinx HBM subsystem has 16 memory channels, 32 pseudo channels, and 32 HBM channels. 

On the top of 16 memory channels, there are 32 \emph{AXI channels} that interact with the user logic. Each AXI channel adheres to the standard AXI3 protocol~\cite{hbm_ip} to provide a proven standardized interface to the FPGA programmer. Each AXI channel is associated with a HBM channel (or pseudo channel), so each AXI channel is only allowed to access its own memory region. To make each AXI channel able to access the full HBM space, Xilinx introduces a switch between 32 AXI channels and 32 pseudo channels~\cite{hbm_wp,hbm_ip}.\footnote{By default, we disable the switch in the HBM memory controller when we measure latency numbers of HBM, since the switch that enables global addressing among HBM channels is not necessary. The switch is on when we measure throughput numbers.} However, the switch is not fully implemented due to its huge resource consumption. Instead, Xilinx presents eight \emph{mini-switches}, where each mini-switch serves four AXI channels and their associated pseudo channels and the mini-switch is fully implemented in a sense that each AXI channel accesses any pseudo channel in the same mini-switch with the same latency and throughput. Besides, there are two bidirectional connections between two adjacent mini-switches for global addressing. 


\vspace{0.5ex}
\section{General Benchmarking Framework \FrameworkName{}}

\vspace{0.5ex}
\subsection{Design Methodology}
We summarize two concrete challenges {\bf C1} and {\bf C2}, and then present \FrameworkName{} to tackle the two challenges.

\vspace{0.5ex}
First, high-level insight ({\bf C1}). 
It is critical to make our benchmarking framework meaningful to FPGA programmers in a sense that we should provide high-level insights to FPGA programmers for ease of understanding. In particular, we should give the programmer an end-to-end explanation, rather than just incomprehensible memory timing parameters like row precharge time $T_{RP}$, so that the insights can be used to improve the use of HBM memory on FPGAs. 

\vspace{0.5ex}
Second, easy to use ({\bf C2}). 
It is difficult to achieve ease of use when benchmarking on FPGAs when a small modification might need to reconfigure the FPGA. 
Therefore, we intend to minimize the reconfiguration effort so that  the FPGA does not need to be reconfigured between benchmarking tasks.  
In other words, our benchmarking framework should allow us to use a single FPGA image for a large number of benchmarking tasks, not just for one benchmarking task. 


\subsubsection{Our Approach}
We propose \FrameworkName{} to tackle the above two challenges. In order to tackle the first challenge {\bf C1}, \FrameworkName{} allows to directly analyze the performance characteristics of typical memory access patterns used by FPGA programmers, providing an end-to-end explanation for the overall performance. 
To tackle the second challenge {\bf C2}, \FrameworkName{} uses runtime parameterization of the benchmarking circuit so as to cover a wide range of benchmarking tasks without reconfiguring the FPGA. Through the access patterns implemented in the benchmark, we are able to unveil the underlying characteristics of HBM and DDR4 on FPGAs. 

\FrameworkName{} adopts a software-hardware co-design approach based on two components: a software component (Subsection~\ref{subsec_software}) and a hardware component (Subsection~\ref{subsec_hardware}). The main role of the software component is to provide flexibility to the FPGA programmer in terms of runtime parameters. With these runtime parameters, we do not need to frequently reconfigure the FPGA when benchmarking HBM and DDR4. The main role of the hardware component is to guarantee performance. More precisely, \FrameworkName{} should be able to expose the performance potential, in terms of maximum achievable memory bandwidth and minimum achievable latency, of HBM memory on the FPGA. To do so, the benchmarking circuit itself cannot be the bottleneck at any time. 



\vspace{-1ex}
\subsection{Software Component}
\label{subsec_software}
\begin{figure}[b]
	\centering
	\vspace{-2ex}
	\includegraphics[width=8.0cm]{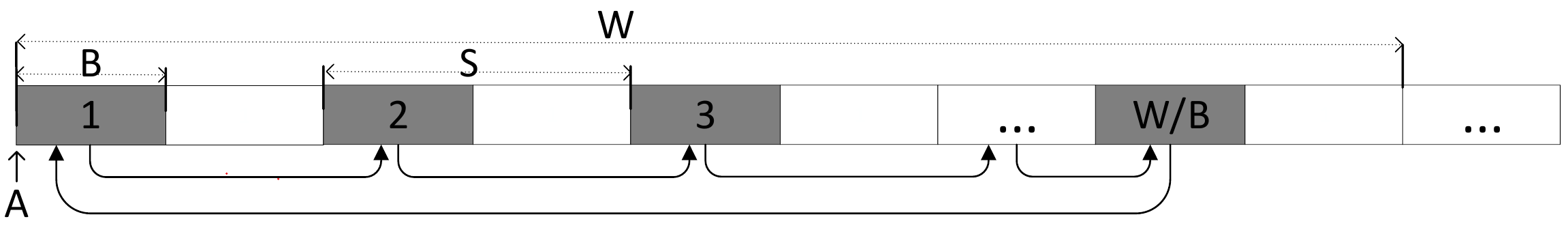}
	\vspace{-1ex}
	\caption{Memory access pattern used in \FrameworkName{}. }
	\vspace{-1ex}
	\label{P_memory_access_pattern}
\end{figure}

\FrameworkName{}'s software component aims to provide a user-friendly interface such that an FPGA developer can easily use \FrameworkName{} to benchmark HBM memory and obtain relevant performance characteristics. To this end, we introduce a memory access pattern widely used in FPGA programming: \emph{Repetitive Sequential Traversal (RST)}, as shown in Figure~\ref{P_memory_access_pattern}. 

The RST pattern traverses a \emph{memory region}, a data array storing data elements in a sequence. 
The RST repetitively sweeps over the memory region of size $W$ with the starting address $A$, and each time reads $B$ bytes with a stride of $S$ bytes, where $B$ and $S$ are a power of 2. On our tested FPGA, the burst size $B$ should be not smaller than 32 (or 64) for HBM (or DDR4) due to the constraint of HBM/DDR4 memory application data width. The stride $S$ should be not larger than the working set size $W$. The parameters are summarized in Table~\ref{t_hardare_platform}. 
We calculate the address $T[i]$ of the $i$-th memory read/write transaction issued by the RST, as illustrated in Equation~\ref{E_addr}. The calculation can be implemented with simple arithmetic, which in turn leads to fewer FPGA resources and potentially higher frequency. Even though the supported memory access pattern is quite simple, it can still unveil the performance characteristics of the memory, e.g., HBM and DDR4, on FPGAs.
\begin{equation} \begin{scriptsize}
\label{E_addr}
T[i] = A + (i \times S) \% W
\end{scriptsize} \end{equation}
\vspace{-1ex}


\begin{table} [t]
	\centering
		\begin{scriptsize}
	\caption{Summary of runtime parameters}
	\vspace{-1ex}
	\label{t_hardare_platform}
	\begin{tabular}{|c||c|}
		\hline
		{\bf Parameter} &  {\bf Definition} \\
		\hline
		\hline
		N & Number of memory read/write transactions\\
		\hline
		B & Burst size (in bytes) of a memory read/write transaction\\
		\hline
		W & Working set size (in bytes). W ($>$16) is a power of 2. \\
		\hline
		S & Stride (in bytes) \\
		\hline
		A & Initial address (in bytes) \\
		\hline
	\end{tabular}
		\end{scriptsize}
\end{table}

\vspace{-1ex}
\subsection{Hardware Component}
\label{subsec_hardware}

\begin{figure}[t]
	\centering
	\includegraphics[width=7.0cm]{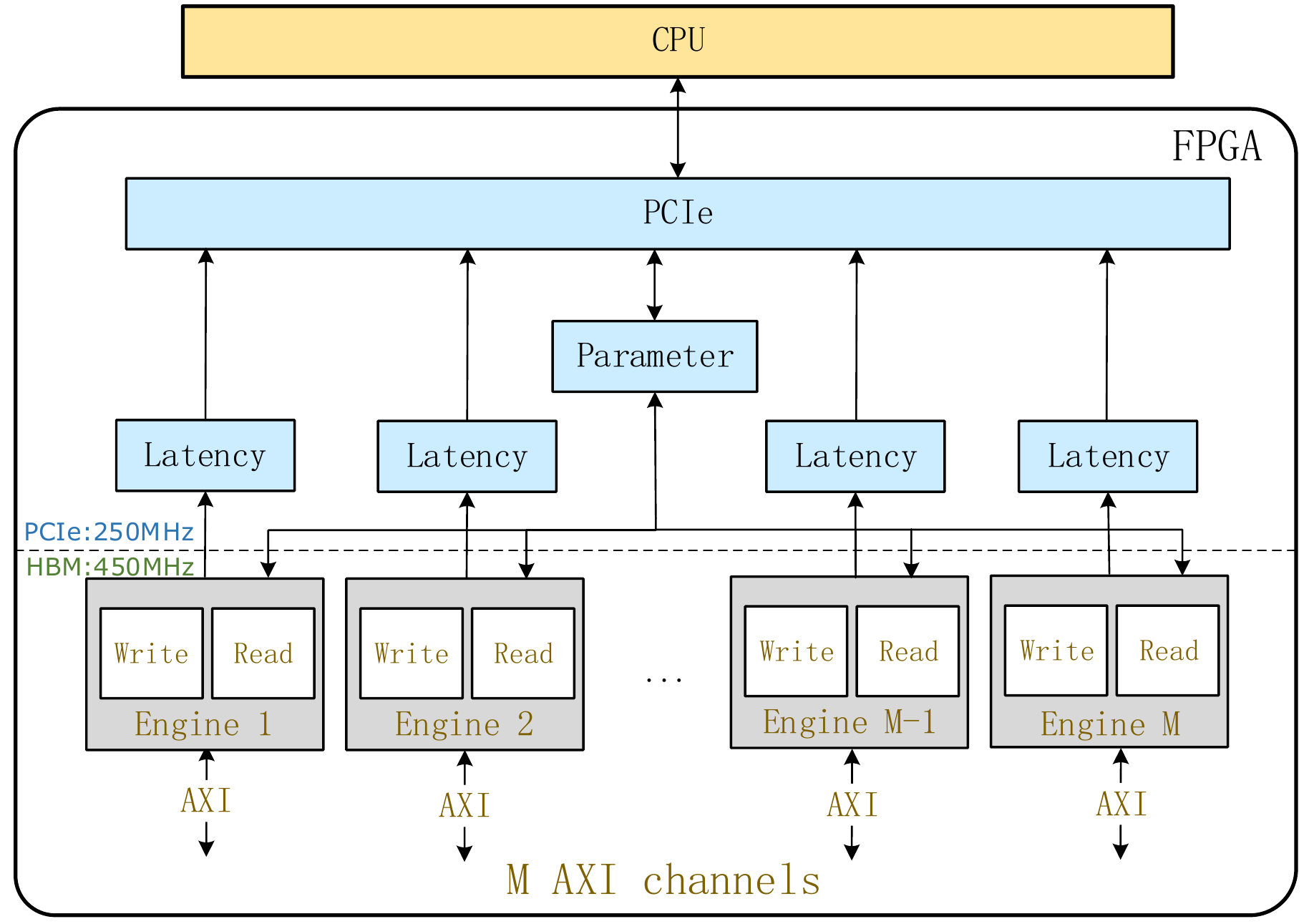}
	\vspace{-1ex}
	\caption{Overall hardware architecture of our benchmarking framework. It can support $M$ hardware engines running simultaneously, with each engine for one AXI channel. In our experiment, $M$ is 32 for HBM, while $M$ is 2 for DDR4. }
	\vspace{-2ex}
	\label{P_overall_hw_architecture}
\end{figure}

The hardware component of \FrameworkName{} consists of a \emph{PCIe module}, $M$ \emph{latency modules}, a \emph{parameter module} and $M$ \emph{engine modules}, as illustrated in Figure~\ref{P_overall_hw_architecture}. In the following, we discuss the implementation details for each module. 

\subsubsection{Engine Module}
We directly attach an instantiated engine module to an AXI channel such that the engine module directly serves the AXI interface, e.g., AXI3 and AXI4~\cite{axi_xilinx_11, axi_specification_17}, provided by the underlying memory IP core, e.g., HBM and DDR4. The AXI interface consists of five different channels: read address (RA), read data (RD), write address (WA), write data (WD) and write response (WR)~\cite{axi_xilinx_11}. Besides, the input clock of the engine module is exactly the clock from the associated AXI channel. For example, the engine module is clocked with 450 MHz when benchmarking HBM as it allows at most 450 MHz for its AXI channels. There are two benefits to use the same clock. First, no extra noise, such as longer latency, is introduced by FIFOs needed to cross different clock regions. Second, the engine module is able to saturate its associated AXI channel, not leading to  underestimates of the memory bandwidth capacity. 

The engine module, written in Verilog, consists of two independent modules: a \emph{write module} and a \emph{read module}. The write module serves three write-related channels WA, WD, and WR, while the read module serves two read-related channels RA and RD.  


\vspace{0.5ex}
The write module contains a state machine to serve a memory-writing task at a time from the CPU. The task has the initial address $A$, number of write transactions $N$, burst size $B$, stride $S$, and working set size $W$. Once the writing task is received, this module always tries to saturate the memory write channels WR and WD by asserting the associated valid signals before the writing task completes, aiming to maximize the achievable throughput. The address of each memory write transaction is specified in Equation~\ref{E_addr}. This module also probes the WR channel to validate that the on-the-fly memory write transactions are successfully finished. 

\vspace{0.5ex}
The read module contains a state machine to serve a memory-reading task at a time from the CPU. The task has the initial address $A$, number of read transactions $N$, burst size $B$, stride $S$, and working set size $W$. Unlike the write module, that only measures the achievable throughput, the read module measures as well the latency of each of serial memory read transactions: we immediately issue the second memory read transaction only after the read data of the first read transaction is returned.\footnote{We are able to unveil many performance characteristics of HBM and DDR4 by analyzing the latency difference among serial memory read transactions. The fundamental reason of the immediate issue is that a refresh command that occurs periodically will close all the banks in our HBM/DDR4 memory, and then there will be no latency difference if the time interval of two serial read transactions is larger than the time (e.g., 7.8 $\mu s$) between two refresh commands. } 
When measuring throughput, this module always tries to saturate the memory read channels RA and RD by always asserting the RA valid signal before the reading task completes. 

\subsubsection{PCIe Module}
We directly deploy the Xilinx DMA/Bridge Subsystem for PCI Express (PCIe) IP core in our \emph{PCIe module}, which is clocked at 250 MHz. Our PCIe kernel driver exposes a PCIe bar mapping the runtime parameters on the FPGA to the user such that the user is able to directly interact with the FPGA using software code. These runtime parameters determine the control and status registers stored in the parameter module. 

\subsubsection{Parameter Module}
The parameter module maintains the runtime parameters and communicates with the host CPU via the PCIe module, receiving the runtime parameters, e.g., $S$, from the CPU and returning the throughput numbers to the CPU. 

Upon receiving runtime parameters, we use them to configure $M$ engine modules, each of which needs two 256-bit control registers to store its runtime parameters: one register for the read module and the other register for the write module in each engine module. Inside a 256-bit register, $W$ takes 32 bits, $S$ takes 32 bits, $N$ takes 64 bits, $B$ takes 32 bits, and $A$ takes 64 bits. The remaining 32 bits are reserved for future use. 
After setting all the engines, the user can trigger the start signal to begin the throughput/latency testing. 

The parameter module is also responsible for returning the throughput numbers (64-bit status registers) to the CPU. One status register is dedicated to each engine module. 

\subsubsection{Latency Module}
We instantiate a \emph{latency module} for each engine module dedicated to an AXI channel. 
The latency module stores a \emph{latency list} of size 1024, where the latency list is written by the associated engine module and read by the CPU. Its size is a \emph{synthesis parameter}. Each latency number containing an 8-bit register refers to the latency for a memory read operation, from the issue of the read operation to the data having arrived from the memory controller.

\vspace{0.5ex}
\section{Experiment Setup}
\vspace{1ex}
\subsection{Hardware Platform}
\label{subsec_platform}
We run our experiments on a Xilinx's Alevo U280~\cite{u280} featuring two HBM stacks of a total size of 8GB and two DDR4 memory channels of a total size of 32 GB. The theoretical HBM memory bandwidth can reach 450 GB/s (450 MHz * 32 * 32 B/s), while the theoretical DDR4 memory bandwidth can reach 38.4 GB/s (300 MHz * 2 * 64 B/s). 
\vspace{-1ex}
\subsection{Address Mapping Policies}
\label{subsec_address_policy}
The application address can be mapped to memory address using multiple policies, where different address bits map to bank, row, or column addresses. Choosing the right mapping policy is critical to maximize the overall memory throughput. 
The  policies enabled for HBM and DDR4 are summarized in Table~\ref{t_address_mapping}, where ``xR" means that $x$ bits are for row address, ``xBG" means that $x$ bits are for bank group address, ``xB" means that $x$ bits are for bank address, and ``xC" means that $x$ bits are for column address. The default policies of HBM and DDR4 are ``RGBCG" and ``RCB", respectively. ``-" stands for  address concatenation. We always use the default memory address mapping policy for both HBM and DDR4 if not particularly specified. For example, the default policy for HBM is RGBCG. 


\begin{table} [t]
	\centering
		\begin{scriptsize}
	\vspace{-1.5ex}
`	\caption{Address mapping policies for HBM and DDR4. The default policies of HBM and DDR4 are marked blue. }
	\label{t_address_mapping}
	\begin{tabular}{|c||c|c|}
		\hline
		{\bf Policies} &  {\bf HBM (app\_addr[27:5])} &  {\bf DDR4 (app\_addr[33:6])}\\
		\hline
		\hline
		{\bf RBC} & 14R-2BG-2B-5C & 17R-2BG-2B-7C\\ 
		\hline
		{\bf RCB} & 14R-5C-2BG-2B & \blue{17R-7C-2B-2BG}\\ 
		\hline
		{\bf BRC} & 2BG-2B-14R-5C & 2BG-2B-17R-7C\\ 
		\hline
		{\bf RGBCG} & \blue{14R-1BG-2B-5C-1BG} &  \\ 
		\hline
		{\bf BRGCG} & 2B-14R-1BG-5C-1BG &  \\ 
		\hline
        {\bf RCBI} & &17R-6C-2B-1C-2BG \\
		\hline
	\end{tabular}
	\vspace{-1ex}

		\end{scriptsize}
\end{table}

\vspace{-1ex}
\subsection{Resource Consumption Breakdown}
\label{subsec_resource_consumption}
In this subsection, we breakdown the resource consumption of the hardware design of \FrameworkName{} when benchmarking HBM.\footnote{Due to space constraints, we omit the resource consumption for benchmarking DDR4 memory on the FPGA.} Table~\ref{t_resource_consumption} shows the exact FPGA resource consumption of each instantiated module. We observe that \FrameworkName{} requires a reasonably small amount of resources to instantiate 32 engine modules, as well as additional components such as the PCIe module, with the total resource utilization being less than 8$\%$.

\begin{table} [t]
	\centering
		\begin{scriptsize}
	\vspace{-0.5ex}
	\caption{Resource consumption breakdown of the hardware design for benchmarking HBM  }
	\label{t_resource_consumption}
	\begin{tabular}{|c||c|c|c|c|c|}
		\hline
		{\bf Hardware modules} &  {\bf LUTs} &  {\bf Registers} &  {\bf BRAMs} &  {\bf Freq.}\\
		\hline
		\hline
		{\bf Engine} & 25824 & 34048 & 0 & 450MHz\\ 
		\hline
		{\bf PCIe} & 70181 & 66689 & 4.36Mb & 250MHz\\ 
		\hline
		{\bf Parameter} & 1607 & 2429 & 0 & 250MHz\\ 
		\hline
		{\bf Latency} & 672 & 1760 & 1.17Mb & 250MHz \\ 
		\hline
		\hline
		{ Total resources used} & 104K  & 122K & 5.53Mb &  \\ 
		\hline
		{Total utilization} & 8$\%$ & 5$\%$ & 8$\%$ &  \\ 
		\hline
	\end{tabular}
	\vspace{-2ex}

		\end{scriptsize}
\end{table}
\vspace{-1ex}
\subsection{Benchmarking Methodology}
\label{subsection_methodology}
We aim to unveil the underlying details of HBM stacks on Xilinx FPGAs under \FrameworkName{}. As a yardstick, we also analyze the performance characteristics of DDR4 on the same FPGA board U280~\cite{u280} when necessary. When we benchmark a HBM channel, we compare the performance characteristics of HBM with that of DDR4 (in Section~\ref{sec_hbm_channel}). We believe that the numbers obtained for a HBM channel can be generalized to other computing devices such as CPUs or GPUs featuring HBMs. When benchmarking the switch inside the HBM memory controller, we do not do the comparison with DDR, since the DDR4 memory controller does not contain such a switch (Section~\ref{sec_switch}).

\vspace{-1ex}
\section{Benchmarking an HBM Channel}
\label{sec_hbm_channel}


\subsection{Effect of Refresh Interval}
\begin{figure}
	\centering
	\subfloat[HBM]{\includegraphics[width=2.3in]{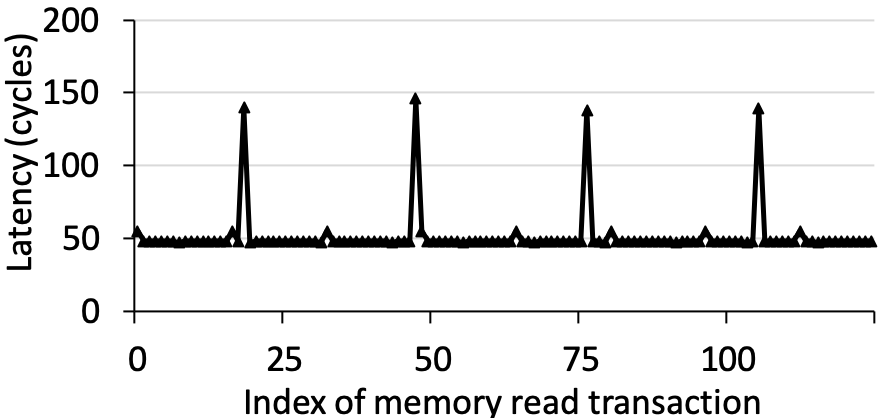} \label{fig_refresh_hbm}} 
    \hfill
	\subfloat[DDR4]{\includegraphics[width=2.3in]{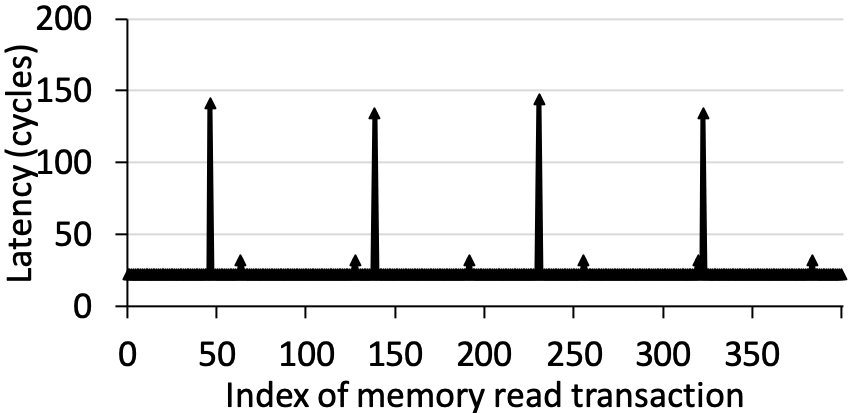} \label{fig_refresh_ddr}} 
	\caption{Higher access latency of memory refresh commands that occur periodically on HBM and DDR4. } 
	\vspace{-2.5ex}
	\label{fig_refresh} 
\end{figure}

When a memory channel is operating, memory cells should be refreshed repetitively such that the information in each memory cell is not lost. During a refresh cycle, normal memory read and write transactions are not allowed to access the memory. We observe that a memory transaction that experiences a memory refresh cycle exhibits a significantly longer latency than a normal memory read/write transaction that is allowed to directly access the memory chips. Thus, we are able to roughly determine the refresh interval by leveraging memory latency differences between normal and in-a-refresh memory transactions. In particular, we leverage \FrameworkName{} to measure the latency of serial memory read operations. Figure~\ref{fig_refresh} illustrates the case with $B$ = 32, $S$ = 64, $W$ = 0x1000000, and $N$ = 1024. We have two observations. First, for both HBM and DDR4, a memory read transaction that coincides with an active refresh command has significantly longer latency, indicating the need to issue enough on-the-fly memory transactions to amortize the negative effect of refresh commands. Second, for both HBM and DDR4, refresh commands are scheduled periodically, the interval between any two consecutive refresh commands being roughly the same.

\subsection{Memory Access Latency}
\label{sub_experiment_mal}
We leverage \FrameworkName{} to accurately measure the latency of consecutive memory read transactions when the memory controller is in an ``idle" state, i.e., where no other pending memory transactions exist in the memory controller such that the memory controller is able to return the requested data to the read transaction with minimum latency. We aim to identify latency cycles of three categories: \emph{page hit}, \emph{page closed}, and \emph{page miss}.\footnote{The latency numbers are identified when the switch is disabled. The latency numbers will be seven cycles higher when the switch is enabled, as the AXI channel accesses its associated HBM channel through the switch. The switching of bank groups does not affect memory  access latency, since at most one memory read transaction is active at any time in this experiment.} 

The ``page hit" state occurs when a memory transaction accesses a row that is open in its bank, so no Precharge and Activate commands are required before the column access, resulting in minimum latency.  

The ``page closed" state occurs when a memory transaction accesses a row whose corresponding bank is closed, so the row Activate command is required before the column access.  

The ``page miss" state occurs when a memory transaction accesses a row that does not match the active row in the bank, so one Precharge command and one Activate command are issued before the column access, resulting in maximum latency.  

We employ the read module to accurately measure the latency numbers for the cases $B$ = 32, $W$ = 0x1000000, $N$ = 1024, and varying $S$. 
Intuitively, the small $S$ leads to high probability to hit the same page while a large $S$ potentially leads to a page miss. Besides, a refresh command closes all the active banks. 
In this experiment, we use two values of $S$: 128 and 128K. 

\begin{figure}
	\centering
	\subfloat[HBM]{\includegraphics[width=2.3in]{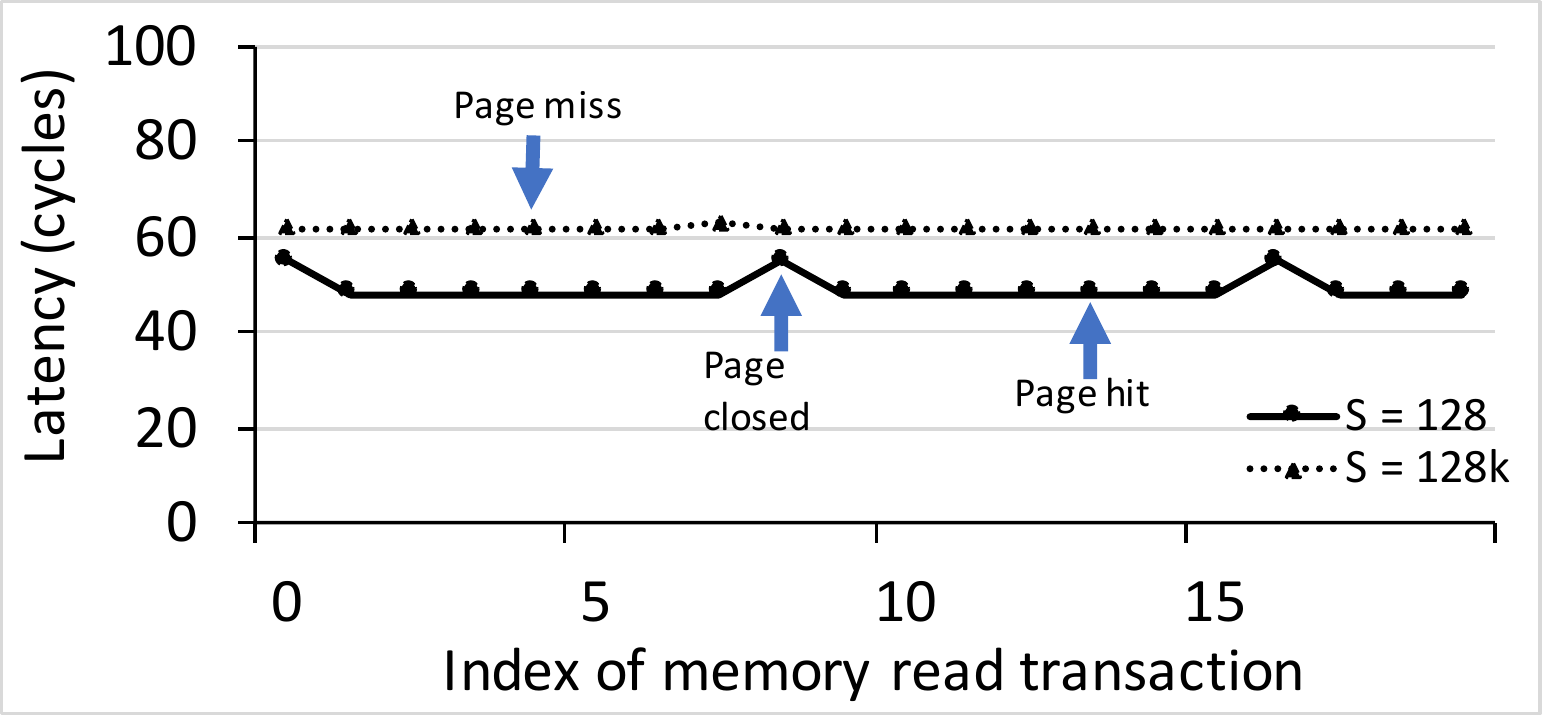} \label{fig_refresh_hbm}} 
    \hfill
	\subfloat[DDR4]{\includegraphics[width=2.3in]{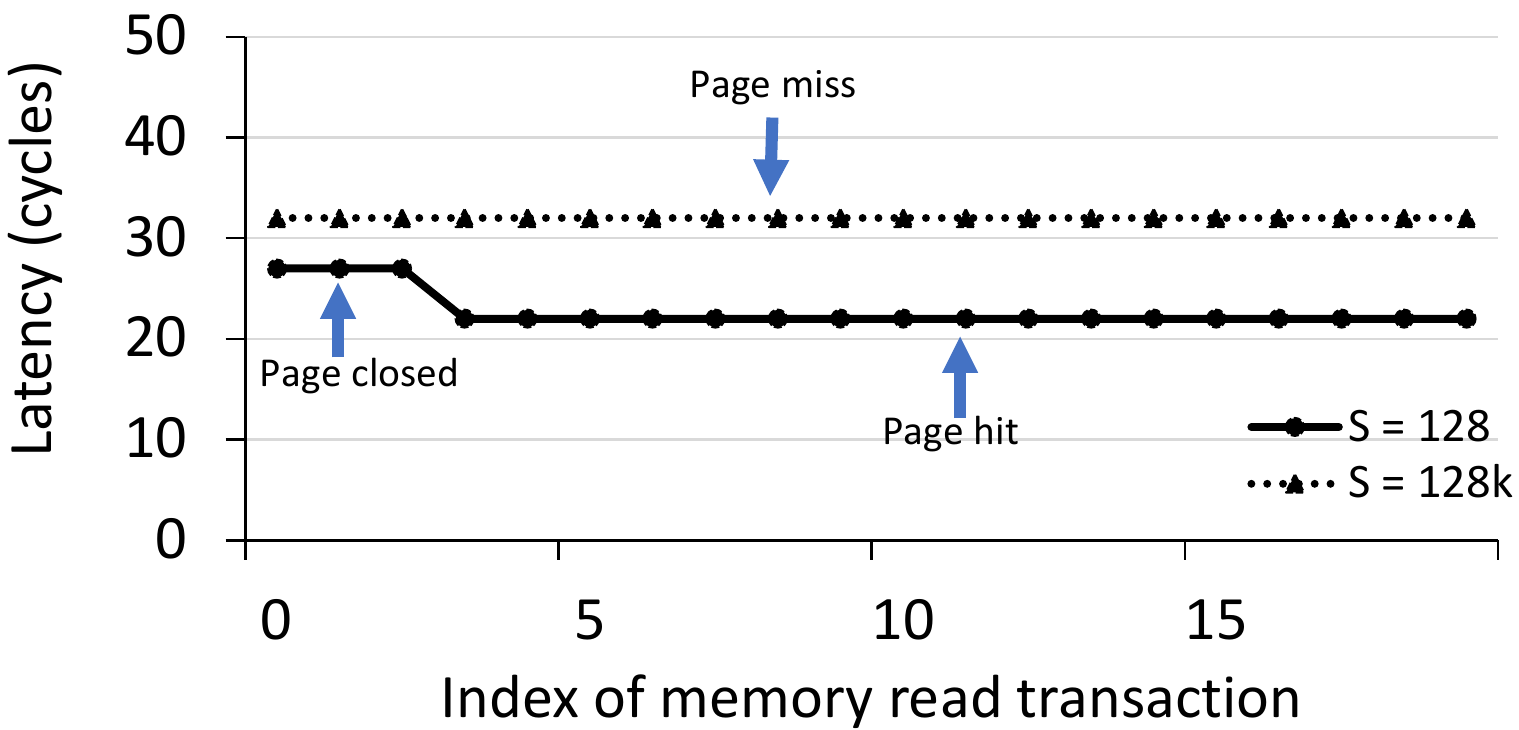} \label{fig_refresh_ddr}} 
	\caption{Snapshots of page miss, page closed and page hit, in terms of latency cycles, on HBM and DDR4. } 
	\vspace{-2ex}
	\label{fig_latency} 
\end{figure}

We use the case $S$=128 to determine the latency of page hit and page closed transactions. $S$=128 is smaller than the page size, so the majority of read transactions will hit an open page, as illustrated in Figure~\ref{fig_latency}. The remaining points illustrate the latency of page closed transactions, since the small $S$ leads to a large amount of read transactions in a certain memory region and then a refresh will close the bank before the access to another page in the same bank.\footnote{The latency trend of HBM is different of that of DDR4 due to the different default address mapping policy. The default address mapping policy of HBM is RGBCG, indicating that only one bank needs to be active at a time, while the default policy of DDR4 is RCB, indicating that four banks are active at a time. } 

We use the case $S$=128K to determine the latency of a page miss transaction. $S$=128K leads to a page miss for each memory transaction for both HBM and DDR4, since two consecutive memory transaction will access the same bank but different pages. 

We summarize the latency on HBM and DDR in Table~\ref{t_idle_latency}. We observe that the memory access latency on HBM is higher than that on DDR4 by about 30 nano seconds under the same category like page hit. It means that HBM could have disadvantages when running latency-sensitive applications on FPGAs.


\begin{table} [t]
	\centering
		\begin{scriptsize}
	\vspace{-0.0ex}
	\caption{Idle memory access latency on HBM and DDR4. Intuitively, the HBM latency is much higher than DDR4. }
	\label{t_idle_latency}
	\begin{tabular}{|c||c|c|c|c|}
		\hline
		    \multirow{2}{*}{{\bf Idle Latency}} & \multicolumn{2}{c|}{{\bf HBM}} & \multicolumn{2}{c|}{{\bf DDR4}} \\
         \cline{2-5}
		& {\bf Cycles} & {\bf Time} & {\bf Cycles} & {\bf Time} \\
		\hline
		\hline
		{\bf Page hit} & 48 & 106.7 ns & 22 & 73.3 ns\\
		\hline
		{\bf Page closed} & 55 & 122.2 ns & 27 & 89.9 ns\\
		\hline
		{\bf Page miss} & 62 & 137.8 ns & 32 & 106.6 ns\\
		\hline

	\end{tabular}
	\vspace{-3.5ex}
		\end{scriptsize}
\end{table}

\subsection{Effect of Address Mapping Policy}
\begin{figure*}
	\centering
	\subfloat[B=32 (HBM)]{\includegraphics[width=2.3in]{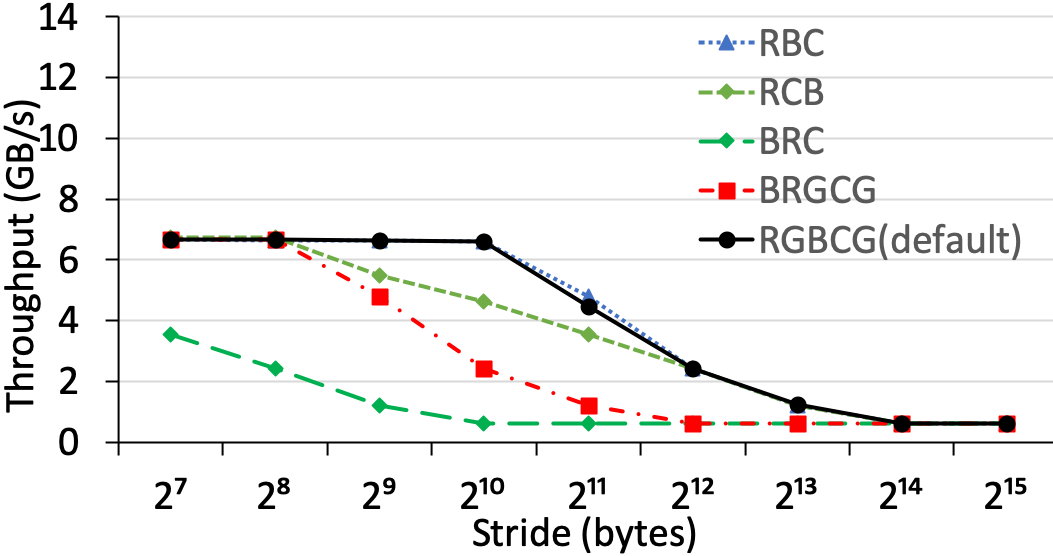} \label{fig_mm_b32}} 
	\subfloat[B=64 (HBM)]{\includegraphics[width=2.3in]{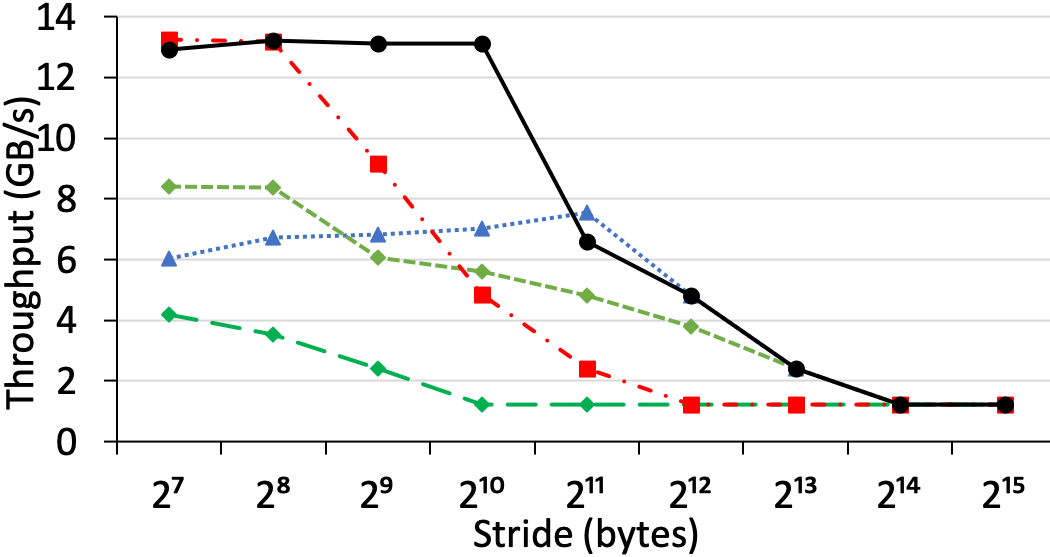} \label{fig_mm_b64}} 
	\subfloat[B=128 (HBM)]{\includegraphics[width=2.3in]{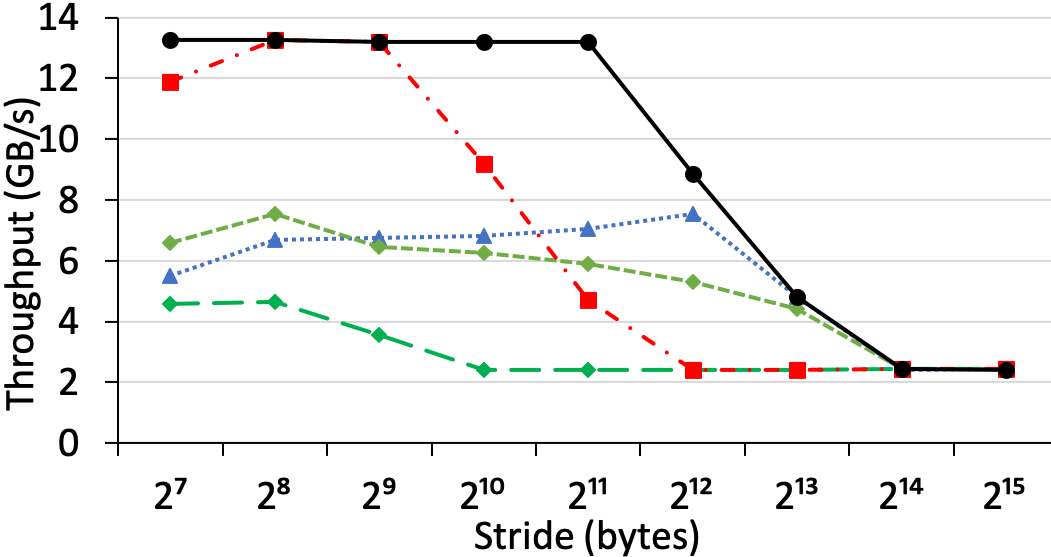} \label{fig_mm_b128}}    \hfill
	\subfloat[B=256 (HBM)]{\includegraphics[width=2.3in]{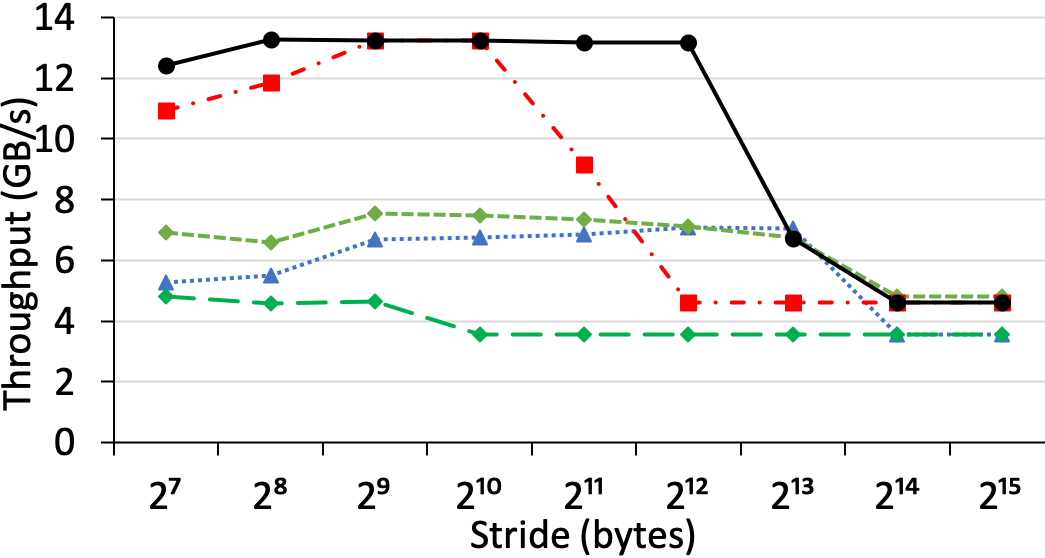} \label{fig_mm_b256}} 	    
	\subfloat[B=64 (DDR4)]{\includegraphics[width=2.3in]{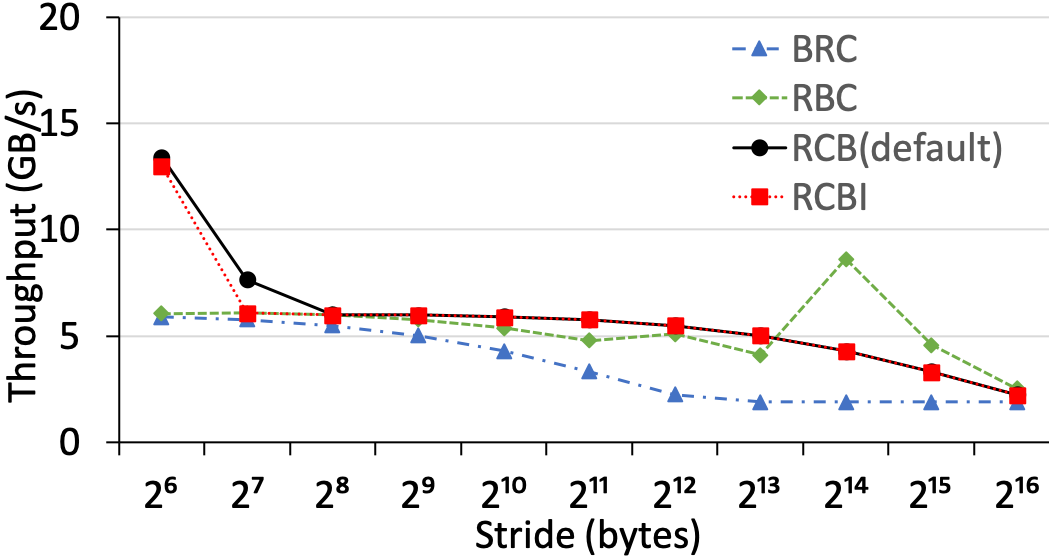} \label{fig_mm_ddr_b64}} 
	\subfloat[B=128 (DDR4)]{\includegraphics[width=2.3in]{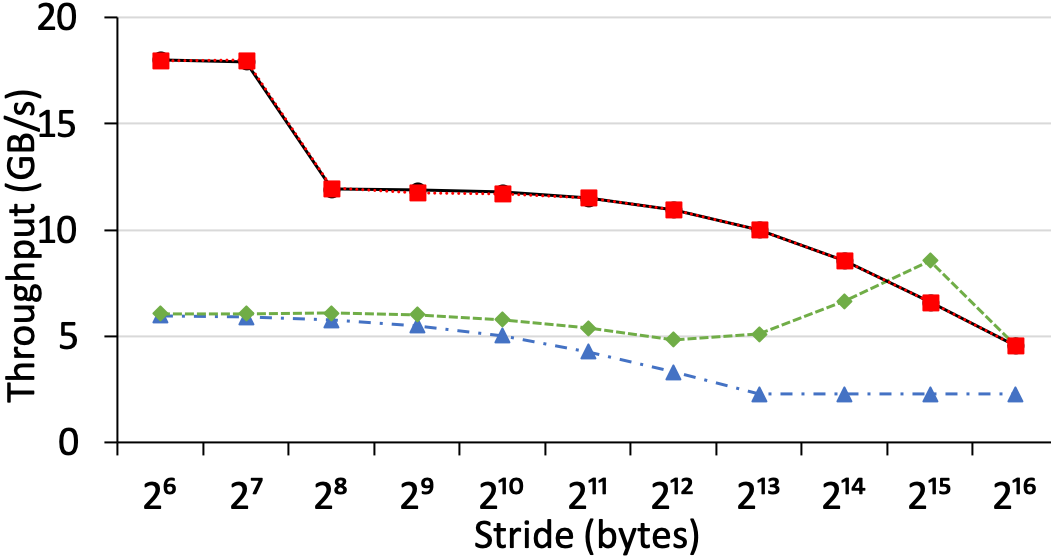} \label{fig_mm_ddr_b128}}     \hfill
	\subfloat[B=256 (DDR4)]{\includegraphics[width=2.3in]{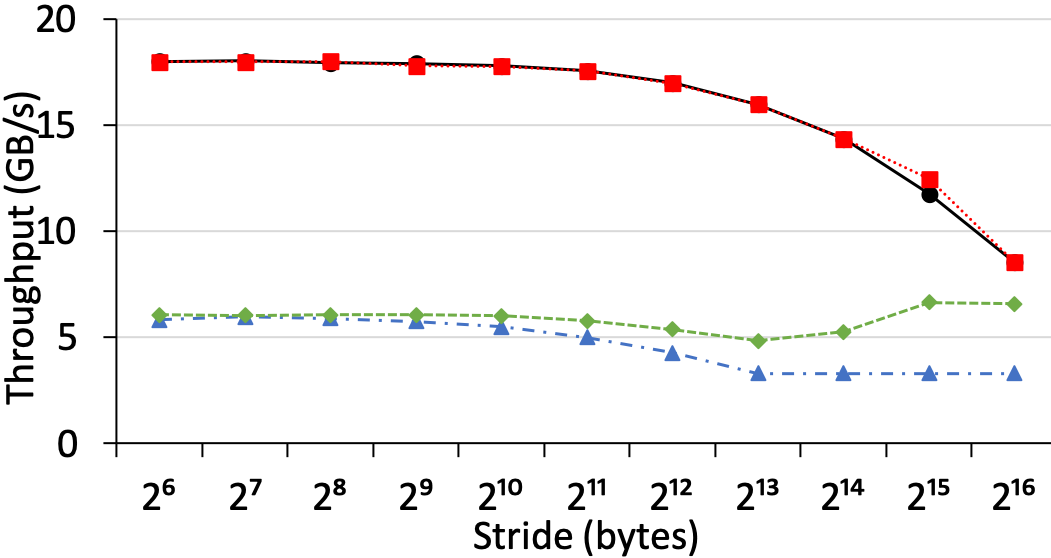} \label{fig_mm_ddr_b256}} 
	\subfloat[B=512 (DDR4)]{\includegraphics[width=2.3in]{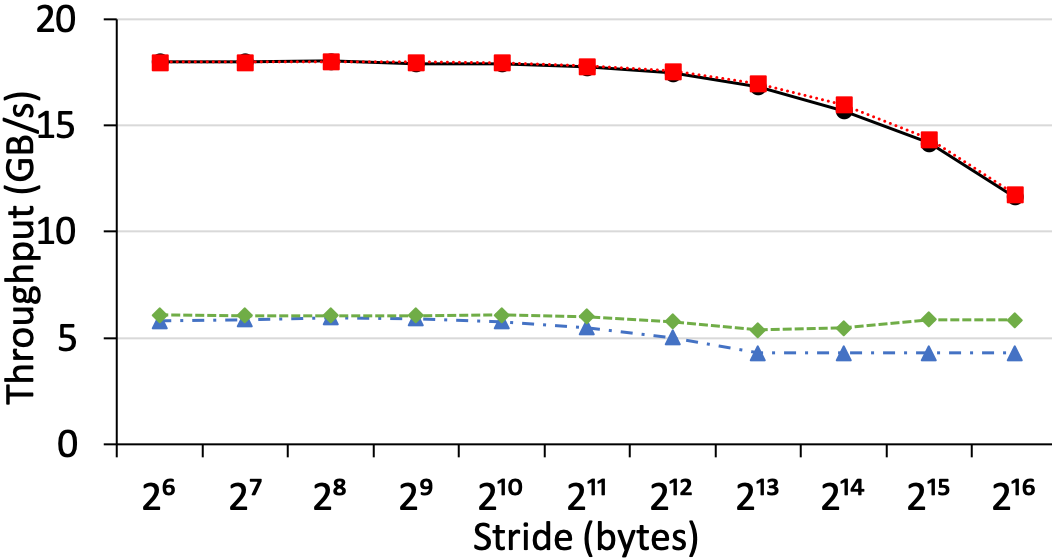} \label{fig_mm_ddr_b512}} 	
	\caption{Memory throughput comparison between an HBM channel and a DDR4 channel, with different burst sizes and stride under all the address mapping policies. In this experiment, we use the AXI channel 0 to access its associated HBM channel 0 for the best performance from a single HBM channel. We use the DDR4 channel 0 to obtain the DDR4 throughput numbers.}
	\vspace{-3ex}
	\label{fig_mm} 
\end{figure*}

In this subsection, we examine the effect of different memory address mapping policies on the achievable throughput. In particular, under different mapping policies, we measure the memory throughput with varying stride $S$ and burst size $B$, while keeping the working set size $W$ (= 0x10000000) large enough. Figure~\ref{fig_mm} illustrates the throughput trend for different address mapping policies for both HBM and DDR4. We have five observations.

First, different address mapping policies lead to significant performance difference. For example, Figure~\ref{fig_mm_b32} illustrates that the default policy (RGBCG) of HBM is almost 10X faster than the policy (BRC) when $S$ is 1024 and $B$ is 32, demonstrating the importance of choosing the right address mapping policy for a memory-bound application running on the FPGA. 
Second, the throughput trends of HBM and DDR4 are quite different even though they employ the same address mapping policy, demonstrating the importance of a benchmark platform such as \FrameworkName{} to evaluate different FPGA boards or different memory generations. 
Third, the default policy always leads to the best performance for any combination of $S$ and $B$ on HBM and DDR4, demonstrating that the default setting is reasonable. 
Fourth, small burst sizes lead to low memory throughput, as shown in Figures~\ref{fig_mm_b32}, \ref{fig_mm_ddr_b64}, meaning that FPGA programmers should increase spatial locality to achieve higher memory throughput out of HBM or DDR4. 
Fifth, large $S$ ($>$8K) always leads to an extremely low memory bandwidth utilization, indicating the extreme importance of keeping spatial locality. In other words, the random memory access that does not keep spatial locality will experience low memory throughput. 
We conclude that choosing the right address mapping policy is critical to optimize memory performance on FPGAs.  

\subsection{Effect of Bank Group}
In this subsection, we examine the effect of bank group, which is a new feature of DDR4, compared to DDR3. Accessing multiple bank groups simultaneously helps us relieve the negative effect of DRAM timing restrictions that have not improved over generations of DRAM. A higher memory throughput can be potentially obtained by accessing multiple bank groups. Therefore, we use the engine module to validate the effect of a bank group (Figure~\ref{fig_mm}). We have two observations. 

First, with the default address mapping policy, HBM allows to use large stride size while still keeping high throughput, as shown in Figures~\ref{fig_mm_b32},~\ref{fig_mm_b64},~\ref{fig_mm_b128},~\ref{fig_mm_b256}. The underlying reason is that even though each row buffer is not fully utilized due to large $S$, bank-group-level parallelism is able to allow us to saturate the available memory bandwidth. 
Second, a pure sequential read does not always lead to the highest throughput under a certain mapping policy. Figures~\ref{fig_mm_b64},~\ref{fig_mm_b128} illustrate that when $S$ increases from 128 to 2048, a bigger $S$ can achieve higher memory throughput under the policy ``RBC", since a bigger $S$ allows more active bank groups to be accessed concurrently, while a smaller $S$ potentially leads to only one active bank group that serves user's memory requests.  
We conclude that it is critical to leverage bank-group-level parallelism to achieve high memory throughput under HBM.

\subsection{Effect of Memory Access Locality}
\begin{figure}
	\centering
	\subfloat[HBM]{\includegraphics[width=2.6in]{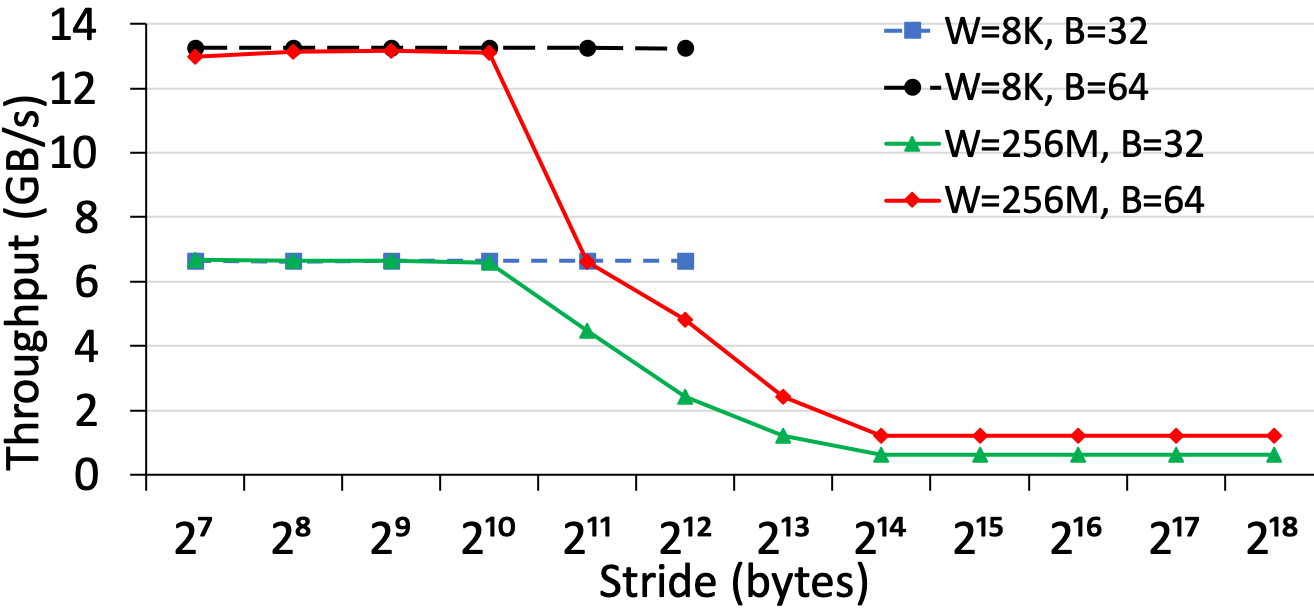} \label{P_locality_hbm}} 
    \hfill
	\subfloat[DDR4]{\includegraphics[width=2.6in]{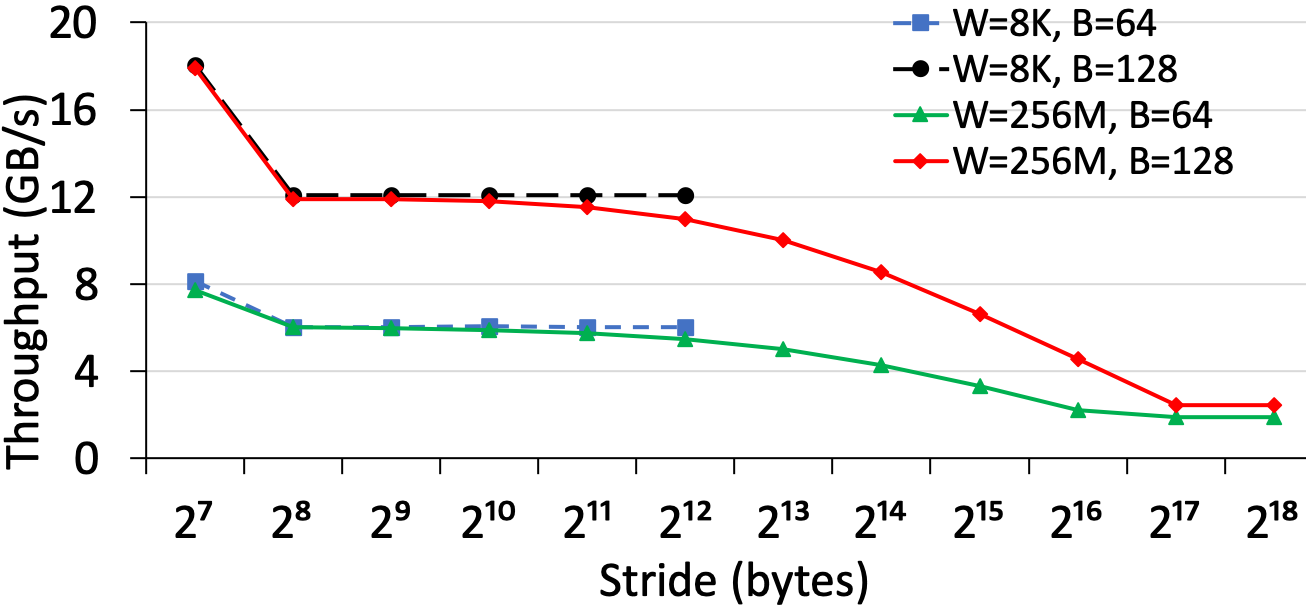} \label{P_locality_ddr}} 
	\caption{Effect of memory access locality.  } 
	\vspace{-2.5ex}
	\label{P_locality} 
\end{figure}

In this subsection, we examine the effect of memory access locality on memory throughput. We vary the burst size $B$ and the stride $S$, and we set the working set size $W$ to two values: 256M and 8K. The case $W$=256M refers to the baseline that does not benefit from any memory access locality, while the case $W$=8K refers to the case that benefits from locality. Figure~\ref{P_locality} illustrates the throughput for varying parameter settings on both HBM and DDR4. We have two observations. 

First, memory access locality indeed increases the memory throughput for each case with high stride $S$. For example, the memory bandwidth of the case ($B$=32, $W$=8K, and $S$=4K) is 6.7 GB/s on HBM, while 2.4 GB/s of the case ($B$=32, $W$=256M, and $S$=4K), indicating that memory access locality is able to eliminate the negative effect of a large stride. Second, memory access locality cannot increase the memory throughput when $S$ is small. In contrast, memory access locality can significantly increase the total throughput on modern CPUs/GPUs due to the on-chip caches which have dramatically higher bandwidth than off-chip memory~\cite{generic_cost_model_Manegold_02}. 


\subsection{Total Memory Throughput}
In this subsection, we explore the total achievable memory throughput of HBM and DDR4 (Table~\ref{t_toatl_throughput}). The HBM system on the tested FPGA card, U280, is able to provide up to 425 GB/s (13.27 GB/s * 32) memory throughput when we use all the 32 AXI channels to simultaneously access their associated HBM channels.\footnote{Each AXI channel accesses its local HBM channel, there is no inference among the 32 AXI channels. Since each AXI channel approximately has the same throughput, we estimate the total throughput by simply scaling up the throughput of the channel 0 by 32.} The DDR4 memory is able to provide up to 36 GB/s (18 GB/s * 2) memory throughput when we simultaneously access both DDR4 channels on our tested FPGA card. We observe that the HBM system has 10 times more memory throughput than DDR4 memory, \wzk{indicating that the HBM-enhanced FPGA enables us to accelerate memory-intensive applications, which are typically accelerated on GPUs}. 
\begin{table} [t]
	\centering
		\begin{scriptsize}
	\vspace{-0.5ex}
	\caption{Total memory throughput comparison between HBM and DDR4.}
	\vspace{-0.5ex}
	\label{t_toatl_throughput}
	\begin{tabular}{|c||c|c|}
		\hline
		{\bf  } &  {\bf HBM} &  {\bf DDR4}\\
		\hline
		\hline
		{\bf Throughput of a channel} & 13.27 GB/s & 18 GB/s\\ 
		\hline
		{\bf Number of channels} & 32 & 2\\ 
		\hline
		{\bf Total memory throughput} & 425 GB/s & 36 GB/s\\ 
		\hline
	\end{tabular}
	\vspace{-3ex}
		\end{scriptsize}
\end{table}



\section{Benchmarking the Switch in the HBM Controller}
\label{sec_switch}

 Our goal in this section is to unveil the performance characteristics of the switch. 
In a fully implemented switch, the performance characteristics of the access from any AXI channel to any HBM channel should be roughly the same. However, in the current implementation, the relative distance could play an important role. In the following, we examine the performance characteristics between any AXI channel and any HBM channel, in terms of latency and throughput. 
 
\subsubsection{Memory Latency}
Due to space constraints, we only demonstrate the memory access latency using the memory read transaction issued in any AXI channel (from 0 to 31) to the HBM channel 0.\footnote{The switch is enabled to allow global addressing, when comparing the latency difference among AXI channels. } Access to other HBM channels has similar performance characteristics. 
Similar to the experimental setup in Subsection~\ref{sub_experiment_mal}, we also employ the engine module to determine the accurate latency for the case $B$ = 32, $W$ = 0x1000000, $N$ = 1024, and varying $S$. Table~\ref{t_latency_32channels} illustrates the latency difference among 32 AXI channels. We have two observations. 

First, the latency difference can be up to 22 cycles. For example, for a page hit transaction, an access from the AXI channel 31 needs 77 cycles, while an access from the AXI channel 0 only needs 55 cycles. Second, the access latency from any AXI channel in the same mini-switch is identical, demonstrating that the mini-switch is fully-implemented. For example, the AXI channels 4-7 in the same mini-switch have the same access latency to the HBM channel 0. 
We conclude that an AXI channel should access its associated HBM channel or the HBM channels close to it to minimize latency. 

\begin{table} [t]
	\centering
		\begin{scriptsize}
	\caption{Memory access latency from any of 32 AXI channels to the HBM channel 0. The switch is on.  Intuitively, longer distance yields longer latency. The latency difference reaches up to 22 cycles.}
	\label{t_latency_32channels}
	\vspace{-1ex}
	\begin{tabular}{|c||c|c|c|c|c|c|}
		\hline
		    \multirow{2}{*}{{\bf Channels }} & \multicolumn{2}{c|}{{\bf Page hit}} &
		    \multicolumn{2}{c|}{{\bf Page closed}} &
		    \multicolumn{2}{c|}{{\bf Page miss}} \\
         \cline{2-7}
		& {\bf Cycles} & {\bf Time} & {\bf Cycles} & {\bf Time} & {\bf Cycles} & {\bf Time} \\
		\hline
		\hline
		{\bf  0-3} & 55 & 122.2 ns & 62 & 137.8 ns & 69 & 153.3 ns\\
		\hline
		{\bf 4-7} & 56 & 124.4 ns & 63 & 140.0 ns & 70 & 155.6 ns\\
		\hline
		{\bf 8-11} & 58 & 128.9 ns & 65 & 144.4 ns & 72 & 160.0 ns\\
		\hline
		{\bf 12-15} & 60 & 133.3 ns & 67 & 148.9 ns & 74 & 164.4 ns\\
		\hline
		{\bf 16-19} & 71 & 157.8 ns & 78 & 173.3 ns & 85 & 188.9 ns\\
		\hline
		{\bf 20-23} & 73 & 162.2 ns & 80 & 177.7 ns & 87 & 193.3 ns\\
		\hline
		{\bf 24-27} & 75 & 166.7 ns & 82 & 182.2 ns & 89 & 197.8 ns\\
		\hline
		{\bf 28-31} & 77 & 171.1 ns & 84 & 186.7 ns & 91 & 202.2 ns\\
		\hline
	\end{tabular}
	\vspace{-1.0ex}
		\end{scriptsize}
\end{table}

\subsubsection{Memory Throughput}
\begin{figure}[t]
	\centering
	\vspace{-1ex}
	\includegraphics[width=7.0cm]{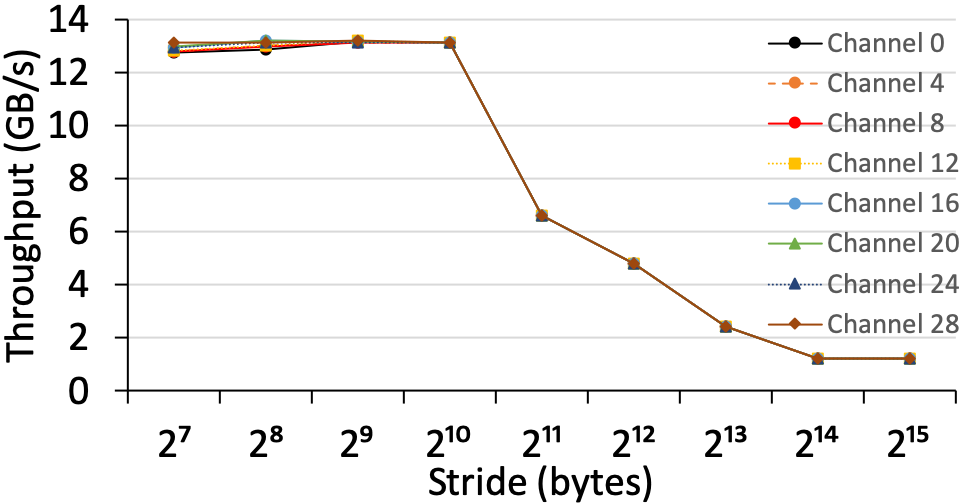}
	\vspace{-1ex}
	\caption{Throughput from eight AXI channels to the HBM channel 1, where each AXI channel is from a mini-switch. }
	\vspace{-3ex}
	\label{P_axi_throughput}
\end{figure}

We employ the engine module to measure memory throughput from any AXI channel (from 0 to 31) to HBM channel 0, with the setting $B$ = 64, $W$ = 0x1000000, $N$ = 200000, and varying $S$. Figure~\ref{P_axi_throughput} illustrates the memory throughput from an AXI channel in each mini-switch to the HBM channel 0. We observe that AXI channels are able to achieve roughly the same memory throughput, regardless of their locations. 

\vspace{0.5ex}
\section{Related Work}
To our knowledge, \FrameworkName{} is the first platform to benchmark HBM on FPGAs in a systematic and comprehensive manner. We contrast closely related work with \FrameworkName{} on 1) benchmarking traditional memory on FPGAs; 2) data processing with HBM; and 3) accelerating application with FPGAs. 

\vspace{0.5ex}
First, benchmarking traditional memory on FPGAs. Previous work~\cite{memory_controller_h2rc19, smartcache_ipdpsw19, memory_analysis_fpt19} tries to benchmark traditional memory, e.g., DDR3, on the FPGA by using high-level languages, e.g., OpenCL. In contrast, we benchmark HBM on the state-of-the-art FPGA.

\vspace{0.5ex}
Second, data processing with HBM/HMC. Previous work~\cite{streambox_asplos19, join_damon18, hashtable_cikm19, dl_sc17, hpc_ipdpsw17, hpc_hbm_sc17, graph_hmc_fpga18, sort_arvix17} employs HBM to accelerate their applications, e.g., hash table deep learning and streaming, by leveraging the high memory bandwidth provided by Intel Knights Landing (KNL)’s HBM~\cite{phi_16}. In contrast, we benchmark the performance of HBM on the Xilinx FPGA. 

\vspace{0.5ex}
Third, accelerating applications with FPGAs. Previous work~\cite{benchamrk_fpga_altera_07, spector_fpl16, benchmark_harp_dac16, benchmark_harp_trts19, fpga_comparison_fpga19, expression_fccm16, ip_fpt18, orb_fpl17, fpgaconvnet_fccm16, affix_fpga19, compressor_tree_dac08, lbw_fpl18, cnn_fpt17, chasing_fpga16, GhostSZ_fccm19, cache_miss_fpga19, network_fpl15, work_stealing_fpga16, smvm_fccm14, genomics_aahpc12, ml_weaving, partition_fpl15, relational_query_processing_fpl16, fpga_opencl_model_hpca16, melia_tpds16, decision_tree_vldb2019, BiSKM_fpga2019, multi_kernel_tvlsi17} accelerates a broad range of applications, e.g., database and deep learning inference, using FPGAs. In contrast, we systematically benchmark HBM on the state-of-the-art FPGA regardless of the application. 

\vspace{0.5ex}
\section{Conclusion}
FPGAs are being enhanced with High Bandwidth Memory (HBM) to tackle the memory bandwidth bottleneck that dominates memory-bound applications. However, the performance characteristics of HBM are still not quantitatively and systematically analyzed on FPGAs. We bridge the gap by benchmarking HBM stack on a state-of-the-art FPGA featuring a two-stack HBM2 subsystem. Accordingly, we propose \FrameworkName{} to demystify the underlying details of HBM such that the user is able to obtain a more accurate picture of the behavior of HBM than what can be obtained by doing so on CPUs/GPUs as they introduce noise from the caches. \FrameworkName{} can be easily generalized to other FPGA boards or other generations of memory modules. We will make the related benchmarking code open-source such that new FPGA boards can be explored and the results across boards are compared. The code is available: https://github.com/RC4ML/Shuhai.

\section*{ Acknowledgements }We thank Xilinx University Program for the valuable feedback to improve the quality of this paper. This work is supported by the National Natural Science Foundation of China (U19B2043, 61976185), and the Fundamental Research Funds for the Central Universities. 

\vspace{0.5ex}

\balance

\bibliographystyle{IEEEtran}
\bibliography{IEEEabrv,myref} 

\end{document}